\documentclass{iopjournal}

\usepackage{amsmath}
\usepackage{cite}
\usepackage{ragged2e}  
\usepackage{microtype}  
\justifying              
\begin{document}

\articletype{Paper} 

\title{Observation of alignment tensor effects in metastability-exchange collisions with highly polarized $^3\mathrm{He}$ ensembles}

\author{Yida Sha$^1$\orcid{0009-0001-5878-2258}, Kaiwen Yi$^1$, Xingqing Jin$^1$, Matteo Fadel$^2$ and Xiang Peng$^{1,*}$}

\affil{$^1$School of Electronics, Peking University, Beijing 100871, China}

\affil{$^2$Department of Physics, ETH Z\"urich, 8093 Z\"urich, Switzerland}

\affil{$^*$Author to whom any correspondence should be addressed.}

\email{xiangpeng@pku.edu.cn}


\begin{abstract}
Highly polarized $^3\mathrm{He}$ ensembles prepared by metastability-exchange optical pumping (MEOP) have been widely used in precision measurements and fundamental physics. 
Metastability-exchange (ME) collisions, serving as the basis of MEOP, are traditionally described in terms of atomic orientation, while the significant contributions of metastable alignment tensor at high polarization remain unexplored.
In this work, we develop a linearized model under mean-field approximation to investigate alignment tensor effects in highly polarized $^3\mathrm{He}$, which originate from the metastable $F=3/2$ manifold and are revealed through ME-induced relaxation and frequency shift.
By means of free-induction-decay (FID) measurements, a pronounced dependence on nuclear polarization is experimentally observed in the response of the ground-state--metastable hybrid $^3\mathrm{He}$ ensembles to the external magnetic field.
Furthermore, after obtaining the characteristics of tensor-induced phenomena, 
we demonstrate good agreement between the experiment and the theory. 
This work advances the understanding of nuclear spin dynamics in highly polarized $^3\mathrm{He}$ using MEOP. It further provides applications in systematic error correction of high-accuracy magnetometry, as well as in optimal protocol for the generation of nuclear spin-squeezed states.
\end{abstract}

\section{Introduction}\label{one}
The ground state of $^3\mathrm{He}$ consists solely of nuclear spin of $I=1/2$, which is well shielded by a closed electronic shell. This results in an extraordinarily long coherence time that can exceed several days~\cite{RevModPhys.89.045004,Schmiedeskamp2006,Deninger2006}. Due to its unique ground-state nuclear spin-1/2 structure and superior coherence, $^3\mathrm{He}$ has found broad applications across various fields, including atomic magnetometers~\cite{3He_LYF,PhysRevLett.124.223001,Koch2015}, biomedical imaging~\cite{EBERT19961297,Qing,Collier_2013zta}, neutron spin filters~\cite{OKUDAIRA2020164301,Zhang2022InSituSEOP,Jiang2014CompactNSF} and fundamental physics~\cite{PhysRevLett.112.110801,PhysRevLett.115.182001,PhysRevLett.129.051802}. 
Recently, quantum memory~\cite{PhysRevA.105.042606,Quantum_memory_3He} and non-classical state schemes~\cite{PhysRevLett.124.043602,PhysRevLett.127.013601,CRPHYS} based on $^3\mathrm{He}$ have also attracted increasing attention.

While isolated nuclear spin is highly immune to environmental disturbances, it also poses challenges for manipulation. Currently, the most effective method for polarizing $^3\mathrm{He}$ is the indirect optical pumping scheme, which can be classified as metastability-exchange optical pumping (MEOP)~\cite{Batz_2011} or spin-exchange optical pumping (SEOP)~\cite{Walker_2011}, depending on the type of optically accessible mediating atoms. MEOP leverages metastable states of $^3\mathrm{He}$ to mediate the coupling through the Coulomb interaction~\cite{PhysRev.132.2561}, whereas SEOP relies on the Fermi-contact interaction with alkali atoms~\cite{PhysRevLett.5.373}. Compared with SEOP, MEOP demonstrates superior nuclear polarization efficiency at low pressure. Additionally, the interaction between ground-state nuclear spins and light can be established and interrupted simply by switching the radio-frequency (RF) discharge, thereby realizing a controllable quantum interface. Conseuqently, MEOP has emerged as a key technique for $^3\mathrm{He}$-based atomic physics studies and quantum sensing applications~\cite{10.1117/12.3074888,PhysRevC.84.022501,RBPartridge_1966,Li_2020,Fadel_2024}.

Metastability-exchange (ME) collisions constitute the mechanism underlying MEOP~\cite{PinardLaloe1980}. Accurate modeling of this process is therefore essential for understanding the indirect optical control of nuclear spins via metastable atoms. The widely used simplified model considers only ME coupling of atomic orientation~\cite{Wang2024}, while ignoring the contributions of the metastable alignment tensor from the $F=3/2$ manifold. This approximation holds for the case of low nuclear polarization (i.e., below 0.25~\cite{DupontRoc1973}). However, with the improvement of pumping efficiency, polarizations above 0.5 have been routinely achieved in MEOP, and values approaching 0.85 can even be reached~\cite{Wang2024,batztel00665393,PANDEY2026170870,refId0,Wu_2026}. According to the spin--temperature distribution~\cite{spin_temperature}, the metastable atoms tend to populate the $m_F=3/2$ Zeeman sublevel at high polarization. Under such conditions, alignment tensors inevitably accumulate; however, their potential effects on ground-state--metastable hybrid $^3\mathrm{He}$ ensembles have not been systematically studied. Such an issue imposes limitations on $^3\mathrm{He}$ MEOP-based quantum technologies, including in high-accuracy magnetometry and the preparation of non-classical states.

In this work, we derive linearized evolution equations for ME collisions that retain tensorial components, which we refer to as the full model.
Our theoretical framework demonstrates that alignment tensors acquire non-negligible stationary solutions, reshaping the ME coupling structure and the associated eigenmodes in highly polarized $^3\mathrm{He}$ ensembles. 
By means of free-induction-decay (FID) measurements,
alignment tensor effects are observed through ME-induced
relaxation and frequency shift. 
Specifically, we quantify the nuclear polarization dependence of ME-induced effects across various external magnetic field strengths, which the simplified model cannot capture.
Our study offers deep insights into the indirect control of $^3\mathrm{He}$ nuclear spins via MEOP.

The remainder of this paper is organized as follows. Section~\ref{two} presents the full model, including the basic theory and the linearization method for ME coupling. In Sec.~\ref{three}, we introduces the experimental setup and the operating principles. Section \ref{four} details verification of the full model via ME-induced relaxation and frequency shift measurements conducted at varying nuclear polarizations. Finally, we give a conclusion in Sec.~\ref{five}. The significance and potential applications of this work are discussed.

\section{Theoretical model}\label{two}
\subsection{Theory}
\begin{figure}[!t]
\centering
\includegraphics[scale=0.7]{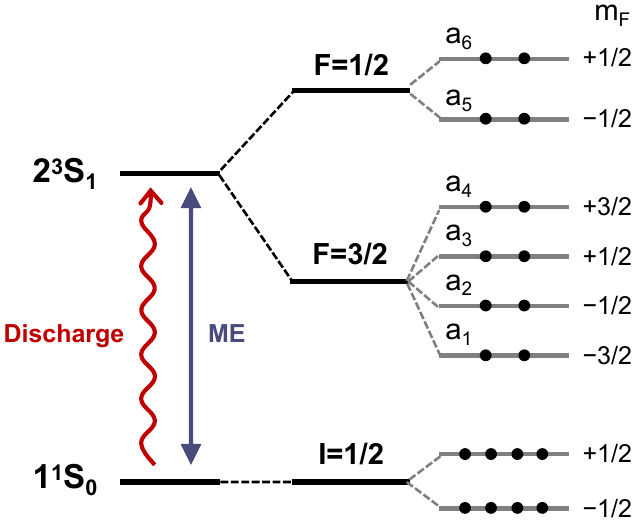}
\caption{$^3\mathrm{He}$ atomic energy-level diagram (not to scale). $1^1\rm S_0$ and $2^3\rm S_1$ represent the ground and metastable states, respectively. The metastable state consists of two hyperfine manifolds with total angular momenta $F = 1/2$ and $F = 3/2$, the corresponding Zeeman sublevel populations being denoted by $a_i$. The RF discharge (red arrow) drives roughly $10^{-6}$ of the total ground-state atoms into the metastable, whereas ME collisions are represented by the blue double arrow.} 
\label{Energylevel}
\end{figure}

The ME coupling involves the ground-state nuclear spin ($\mathbf{I}$) and the metastable manifolds ($\mathbf{F}_{3/2}$ and $\mathbf{F}_{1/2}$), as illustrated by the energy-level structure of $^3\mathrm{He}$ in Fig.~\ref{Energylevel}. The evolution equations for the collective angular momentum variables are expressed as follows~\cite{Fadel_2024}:
\begin{equation}
    \begin{aligned}
        \left.\frac{\mathrm{d}}{\mathrm{d}t}\langle\mathbf{I}\rangle\right|_\text{ME}
        &=- \frac{1}{T_e}\langle\mathbf{I}\rangle+\frac{1}{3T_e}\frac{N_\mathrm{t}}{n_\mathrm{t}}\left(\langle\mathbf{F}_{3/2}\rangle-\langle\mathbf{F}_{1/2}\rangle\right),\\
        \left.\frac{\mathrm{d}}{\mathrm{d}t}\langle\mathbf{F}_{3/2}\rangle\right|_\text{ME}
        &=-\frac{4}{9\tau_e}\langle\mathbf{F}_{3/2}\rangle+\frac{10}{9\tau_e}\langle\mathbf{F}_{1/2}\rangle+\frac{10}{9\tau_e}\frac{n_\mathrm{t}}{N_\mathrm{t}}\langle\mathbf{I}\rangle+\frac{4}{3\tau_e}\frac{1}{N_t}\langle\mathbf{ Q}\rangle\langle\mathbf{I}\rangle,\\
        \left.\frac{\mathrm{d}}{\mathrm{d}t}\langle\mathbf{F}_{1/2}\rangle\right|_\text{ME}
        &=-\frac{7}{9\tau_e}\langle\mathbf{F}_{1/2}\rangle+\frac{1}{9\tau_e}\langle\mathbf{F}_{3/2}\rangle-\frac{1}{9\tau_e}\frac{n_\mathrm{t}}{N_\mathrm{t}}\langle\mathbf{I}\rangle-\frac{4}{3\tau_e}\frac{1}{N_\mathrm{t}}\langle\mathbf{Q}\rangle\langle\mathbf{I}\rangle,\\
        \left.\frac{\mathrm{d}}{\mathrm{d}t}\langle Q_{\alpha\beta}\rangle\right|_\text{ME}
		&=-\frac{2}{3\tau_e}\langle Q_{\alpha\beta}\rangle+\frac{1}{9\tau_e}\frac{1}{N_\mathrm{t}}\times\left[\frac{3}{2}\left(\langle I_\alpha\rangle\langle\Sigma_\beta\rangle+\langle I_\beta\rangle\langle\Sigma_\alpha\rangle\right)-\mathbf{I}\cdot\boldsymbol{\Sigma}\delta_{\alpha\beta}\right].\label{MEC}
    \end{aligned}
\end{equation}
$N_\mathrm{t}$ and $n_\mathrm{t}$ denote the total number of atoms in the ground and metastable states respectively, while $1/T_e$ and $1/\tau_e$ represent the corresponding ME collision rates, satisfying $T_e/\tau_e = N_\mathrm{t}/n_\mathrm{t}$. Specifically, ME collision rates are characterized by the atomic densities and ME rate coefficient $k_\text{ME}\approx154\times10^{-12}\,\mathrm{cm}^3/\text{s}$ at $T=300\,\mathrm{K}$~\cite{RevModPhys.89.045004}. At a gas cell pressure of 86\,Pa, typical values are: $1/T_e\sim 10\,\mathrm{s}^{-1}$ and $1/\tau_e\sim 10^{6}\,\mathrm{s}^{-1}$. The electron spin operator is defined as $\boldsymbol{\Sigma}=\frac23\left(\mathbf{F}_{3/2}+2\mathbf{F}_{1/2}\right)$, while the collective alignment tensor described by Cartesian coordinates is given by $Q_{\alpha\beta}=\frac{1}{18}\left[\frac{3}{2}\left(F_{3/2,\alpha}F_{3/2,\beta}+F_{3/2,\beta}F_{3/2,\alpha}\right)-\mathbf{F}^2_{3/2}\delta_{\alpha\beta}\right]$, where $\alpha, \beta\in\{x,y,z\}$~\cite{DupontRoc1973}. There are twelve tensors in total for the metastable $F=3/2$ manifold, comprising five rank-2 tensors (alignment) and seven rank-3 tensors. Given that the rank-3 tensors do not contribute to the evolution of the orientation and alignment, they are omitted from the subsequent analysis. To facilitate the treatment of the alignment tensor dynamics under ME collisions and an external magnetic field, we adopt the irreducible spherical representation $T^k_q$, satisfying
\begin{equation}
    \begin{gathered}
    Q_{xx}=\frac{1}{6}\left(\sqrt{3}\Re T_2^2-T_0^2\right),\\
    Q_{yy}=-\frac{1}{6}\left(\sqrt{3}\Re T_2^2+T_0^2\right),\\
    Q_{zz}=\frac{1}{3} T_0^2,\\ 
    Q_{xy}=\frac{1}{2\sqrt3}\Im T_2^2,\\ 
    Q_{yz}=-\frac{1}{2\sqrt3}\Im T_1^2,\\
    Q_{xz}=-\frac{1}{2\sqrt3}\Re T_1^2.\label{basis_relation}
\end{gathered}
\end{equation}
The symmetric and antisymmetric combinations $\Re T_q^2=\frac{1}{\sqrt{2}}\left[T_q^2+\left(T_q^2\right)^\dagger\right]$ and $\Im T_q^2=\frac{1}{i\sqrt{2}}\left[T_q^2-\left(T_q^2\right)^\dagger\right]$ refer to the real and imaginary parts of $T_q^2$. 

In ME-coupled $^3\mathrm{He}$ ensembles, the external static magnetic field $\mathbf{B_0}$ is a crucial factor responsible for the relaxation and frequency shift, which arise from the large mismatch between the gyromagnetic ratios of the metastable and ground states. The Hamiltonian $H_B$ is written as
\begin{equation}
        H_{B}=-\hbar\left(\frac{2}{3}\gamma_\mathrm{m}\mathbf{F}_{3/2}\cdot\mathbf{B_0}+\frac{4}{3}\gamma_\mathrm{m}\mathbf{F}_{1/2}\cdot\mathbf{B_0}+\gamma_\text{n}\mathbf{I}\cdot\mathbf{B_0}\right),\label{Hamiltonian}
\end{equation}
where $\gamma_\mathrm{m} = -2\pi\times28.02\,\mathrm{Hz/nT}$ and $\gamma_\mathrm{n} = -2\pi\times32.43\,\mathrm{mHz/nT}$ are the gyromagnetic ratios for electron and nuclear spins~\cite{DupontRoc1973}. According to Heisenberg equation,
\begin{equation}
    \left.\frac{\mathrm{d}}{\mathrm dt}\mathbf{F}\right|_{B}=\frac{i}{\hbar}\left[H_{B},\mathbf{F}\right],
\end{equation}
we obtain the evolution of atomic variables $\mathbf{F}$ (including orientation and alignment) under the external field. More details are provided in \ref{appendixA}.

Regarding the decoherence of $^3\mathrm{He}$, the intrinsic relaxation rate of the ground state, $\Gamma_\mathrm{g}$, is approximately $10^{-2}~\mathrm{s}^{-1}$~\cite{spin_temperature}, while that of the metastable state, $\Gamma_\mathrm{m}$, is about $10^{4}~\mathrm{s}^{-1}$~\cite{Batz_2011}. Compared to the corresponding ME collision rates, intrinsic relaxation contributions are negligible and are therefore omitted from the theoretical model. Consequently, in the presence of both the RF discharge and the external magnetic field, the evolution of the atomic variables is governed by two processes,
\begin{equation}
    \frac{\mathrm{d}}{\mathrm dt}\mathbf{F}=\left.\frac{\mathrm{d}}{\mathrm dt}\mathbf{F}\right|_\text{ME}+\left.\frac{\mathrm{d}}{\mathrm dt}\mathbf{F}\right|_{B}.
\end{equation}

\subsection{Linearization method}
Because of the presence of the alignment tensor, the ME coupling in an external magnetic field takes a nonlinear form. We decompose the operators into a stationary value and a small fluctuation, $\mathbf{F} = \langle\mathbf{F} \rangle_\mathrm{st} + \delta\mathbf{F}$, where $\delta\mathbf{F}$ can be realized by a slight tilt of the collective nuclear spin about the polarization axis~\cite{PhysRevLett.127.013601,Fadel_2024}. By employing the mean-field approximation to neglect higher-order correlations, a set of linearized equations for the fluctuations can be derived, with our analysis focusing on the transverse components. 

\begin{figure}[!t]
\centering
\includegraphics[scale=0.85]{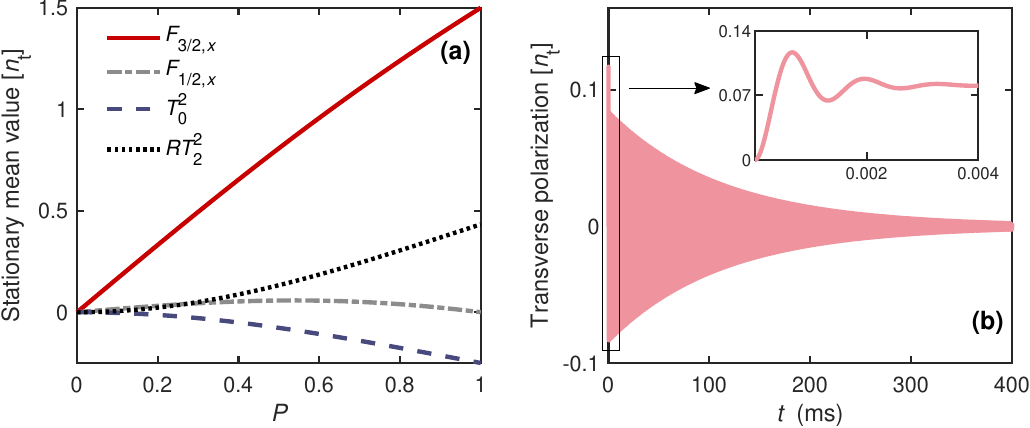}
\caption{(a) Dependence of stationary mean values on nuclear polarization $P$. (b) Time dependence of transverse polarization $\delta F_{3/2,z}$ obtained from the full model. The inset depicts the initial dynamical evolution over the time interval 0 to 0.004~ms. All vertical axes show values normalized to metastable atom number $n_\mathrm{t}$. Simulation parameters in panel (b) are chosen as $B_0=40~\mu\mathrm{T}$, $\tau_e/T_e= 3\times10^{-6}$, $T=300~\mathrm{K}$, $P=0.6$ with the collective nuclear spin tilted by $7^\circ$.} 
\label{Theoretical}
\end{figure}

We assume that the external magnetic field and the collective nuclear spin are aligned along the $x$-axis. This suggests that the stationary mean values are $\langle I_x\rangle_\mathrm{st}=PN_\mathrm{t}/2$ and $\langle I_{y,z}\rangle_\mathrm{st}=0$, where $P$ represents nuclear polarization. Regarding metastable atoms, ME collisions enforce a spin--temperature distribution in angular momentum space. Accordingly, the metastable Zeeman sublevel populations are distributed exponentially over the angular momentum projection, with a Boltzmann parameter defined as $e^\beta=\left(1-P\right)/\left(1+P\right)$, where $1/\beta$ plays the role
of a spin temperature. Therefore, one may write for populations $a_i^\text{st}=n_\mathrm{t}e^{-\beta m_F}/\left(e^{3\beta/2}+2e^{\beta/2}+2e^{-\beta/2}+e^{-3\beta/2}\right)$~\cite{PhysRev.132.2561}. Using the stationary populations $a_i^\mathrm{st}$ in the metastable hyperfine Zeeman sublevels together with their corresponding magnetic quantum numbers $m_F$, we obtain:
\begin{equation}
    \begin{gathered}
       \langle F_{3/2,x}\rangle_\text{st}=P\left(\frac{5+P^2}{3+P^2}\right)n_\mathrm{t},\\ 
       \langle F_{1/2,x}\rangle_\text{st}=\frac{P}{2}\left(\frac{1-P^2}{3+P^2}\right)n_\mathrm{t},\\
       \langle T_0^2\rangle_\text{st}=-\left(\frac{P^2}{3+P^2}\right)n_\mathrm{t},\\
       \langle\Re T_2^2\rangle_\text{st}=\sqrt{3}P\left(\frac{P^2}{3+P^2}\right)n_\mathrm{t}.\label{stationary solutions}
    \end{gathered}
\end{equation}
The transverse components of the metastable orientation vanish in the steady state, as do the other rank-2 alignment tensors $\Re T_1^2$, $\Im T_1^2$, and $\Im T_2^2$. The stationary mean values given by (\ref{stationary solutions}) are plotted in Fig.~\ref{Theoretical}(a). In the low-polarization regime, the contributions of the alignment tensor $T_0^2$ and $\Re T_2^2$ can be ignored, as reported in the previous study~\cite{DupontRoc1973}. Nevertheless, with increasing polarization, the stationary values of alignment tensors become significant, resulting in the modifications of the ME coupling structure that are not captured by the simplified model.

Using the steady-state values derived above and truncating the fluctuations at first order, the ME-coupled equations take the form $\dot{\mathbf{X}}=\mathbf{M}\mathbf{X}$, where $\mathbf{X}=\left(\mathbf{X}_1,\,\mathbf{X}_2\right)$ denotes a 14-dimensional fluctuation vector. $\mathbf{X}_1$ and $\mathbf{X}_2$ represent the longitudinal and transverse components, which are written as
\begin{equation}
    \begin{split}
        &\mathbf{X}_1
        =\big(\delta I_x,\,\delta F_{3/2,x},\,\delta F_{1/2,x},\,\delta\Re T_2^2,\,\delta\Im T_1^2,\,\delta T_0^2\big),\\
        &\hat{\mathbf{X}}_2
        =\big(\delta I_y,\,\delta I_z,\,\delta F_{3/2,y},\,\delta F_{3/2,z},\,\delta F_{1/2,y},\,
        \delta F_{1/2,z},\,\delta\Im T_2^2,\,\delta\Re T_1^2\big).
    \end{split}
\end{equation}
The coupling matrix $\mathbf{M}$ forms a block-diagonal structure, 
\begin{equation}
    \mathbf{M}=
    \begin{pmatrix}
        \mathbf{M}_{11} & 0 \\
        0 & \mathbf{M}_{22},
    \end{pmatrix},
\end{equation}
indicating that the longitudinal and transverse fluctuations evolve in two independent subspaces. Given that our experimental focus is the evolution of the transverse components, only the explicit form of the transverse coupling matrix $\mathbf{M}_{22}$ is shown below, 
\begin{equation}
    \renewcommand{\arraystretch}{1.7} 
    \mathbf{M}_{22}=
    \begin{pmatrix}
            \begin{array}{@{\hspace{1em}}c @{\hspace{1em}} 
               c @{\hspace{1em}} 
               c @{\hspace{1em}} 
               c @{\hspace{1em}} 
               c @{\hspace{1em}}
               c @{\hspace{1em}} 
               c @{\hspace{1em}} 
               c @{\hspace{1em}}}
            -\frac{1}{T_e} & -\omega_\mathrm{n} & \frac{1}{3\tau_e} & 0 & -\frac{1}{3\tau_e} & 0 & 0 & 0\\
            \omega_\mathrm{n} & -\frac{1}{T_e} & 0 & \frac{1}{3\tau_e} & 0 & -\frac{1}{3\tau_e} & 0 & 0\\
            \frac{2\left(5+P^2\right)}{3T_e\left(3+P^2\right)} & 0 & -\frac{4}{9\tau_e} & -\omega_{3/2} & \frac{10}{9\tau_e} & 0 & 0 & \frac{P}{3\sqrt{3}}\\
            0 & \frac{2\left(5+P^2\right)}{3T_e\left(3+P^2\right)} & \omega_{3/2} & -\frac{4}{9\tau_e} & 0 & \frac{10}{9\tau_e} & -\frac{P}{3\sqrt{3}\tau_e} & 0\\
            -\frac{1-P^2}{3T_e\left(3+P^2\right)} & 0 & \frac{1}{9\tau_e} & 0 & -\frac{7}{9\tau_e} & -\omega_{1/2} & 0 & -\frac{P}{3\sqrt{3}\tau_e}\\
            0 & -\frac{1-P^2}{3T_e\left(3+P^2\right)} & 0 & \frac{1}{9\tau_e} & \omega_{1/2} & -\frac{7}{9\tau_e} & \frac{P}{3\sqrt{3}\tau_e} & 0\\
            \frac{4P}{\sqrt{3}T_e\left(3+P^2\right)} & 0 & \frac{P}{3\sqrt{3}\tau_e} & 0 & \frac{2P}{3\sqrt{3}\tau_e} & 0 & -\frac{2}{3\tau_e} & \omega_{3/2}\\
            0 & -\frac{4P}{\sqrt{3}T_e\left(3+P^2\right)} & 0 & -\frac{P}{3\sqrt{3}\tau_e} & 0 & -\frac{2P}{3\sqrt{3}\tau_e} & -\omega_{3/2}& -\frac{2}{3\tau_e}
            \end{array}
    \end{pmatrix},\label{M22}
\end{equation}
while the longitudinal block $\mathbf{M}_{11}$ is provided in \ref{appendixB}. $\omega_{1/2}$, $\omega_{3/2}$, and $\omega_\mathrm{n}$ denote the Larmor frequencies under an external field set along the $x$-axis. According to (\ref{M22}), the eigenmodes of the $^3\mathrm{He}$ ensembles under ME collisions exhibit a critical dependence on the nuclear polarization. While the full model reduces to the simplified model in the low-polarization regime,  as shown in \ref{appendixC}, under highly polarized conditions, the transverse block $\mathbf{M}_{22}$ is markedly reshaped by the alignment tensor. The resulting eigenspectrum comprises eight ME-coupled modes, grouped into two distinct lifetime classes. The first contains two degenerate slow-decay modes dominated by $\delta I_y$ and $\delta I_z$, the lifetimes of which depend on the external field and range from $T_e \sim10^{-1}~\mathrm{s}$ to $1/\Gamma_\mathrm{g}\sim10~\mathrm{s}$~\cite{DupontRoc1973}. The remaining six modes belong to the fast-decay class and are governed by metastable variables, exhibiting characteristic lifetimes of $\tau_e \sim 10^{-6}~\mathrm{s}$~\cite{DupontRoc1973}. 
As illustrated in Fig.~\ref{Theoretical}(b), the fast modes initially dominate the transverse polarization dynamics, yet they decay rapidly. Subsequently, the metastable states adiabatically follow the slowly evolving ground states, enabling optical detection of the nuclear spins. In the following discussion, we focus on the eigenvalue of these slow modes to delineate the effects of the alignment tensor on the ME process.

\section{Experiment setup}\label{three}
\begin{figure}[!t]
\centering
\includegraphics[scale=0.47]{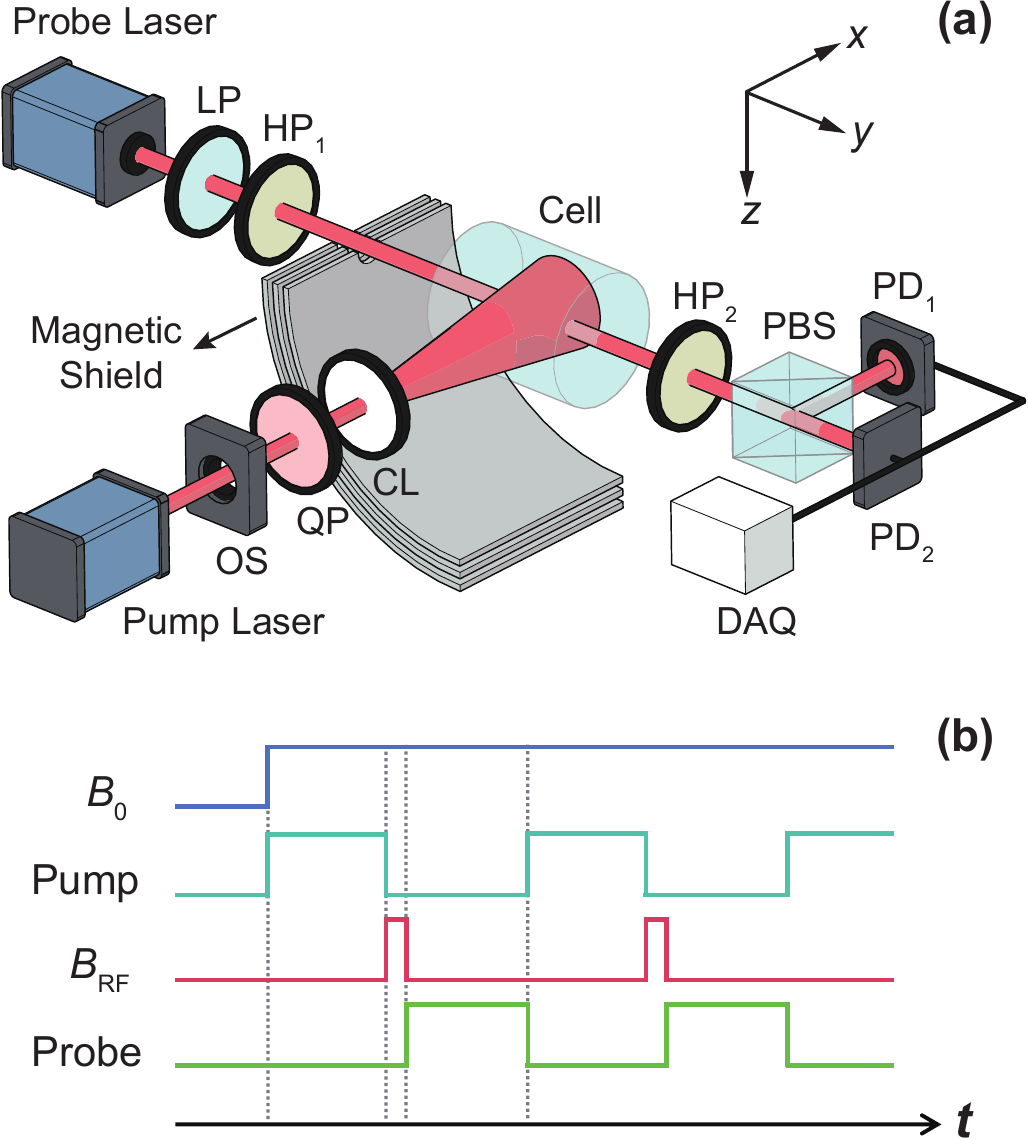}
\caption{(a) Sketch of the experiment apparatus. LP, linear polarizer; HP, half-wave plate; QP, quarter-wave plate; CL, concave lens; PBS, polarizing beam splitter; PD, photoelectric detector; DAQ, data acquisition; OS, optical shutter. (b) Timing diagram for static magnetic field $B_0$, pump laser, RF pulse $B_\mathrm{RF}$ and probe laser. High (low) level corresponds to the ON (OFF) state. } 
\label{Figure3}
\end{figure}

The configuration of the experimental apparatus is depicted in Fig.~\ref{Figure3}(a). We choose a cylindrical cell, 50~mm in diameter and 70~mm in length, filled with pure $^3\mathrm{He}$ gas at 86~Pa. An RF discharge at 32.8~MHz with a power of 0.15~W is used to generate metastable atoms. To suppress magnetic noise, the cell is enclosed at the center of a seven-layer $\mu$-metal magnetic shield. A three-axis Helmholtz coil is used to generate mutually orthogonal fields with a static magnetic field provided by a DC current source (Krohn-Hite Model 523), while the RF pulse is driven by a waveform generator (Keysight 33500B). 
The pump light resonant with the $C_9$ transition~\cite{spin_temperature} is emitted from a fiber laser (NKT Koheras BASIK Y10). An optical shutter (OS) controls the temporal sequence of the pump light. Before propagating into the cell, the pump beam is converted to circular polarization and expanded to approximately 50 mm in diameter by a quarter-wave plate and a concave lens. We employ a distributed Bragg reflector (DBR) diode laser (Photodigm PH1083DBR040BF-ISO) in the probe path, operating at a wavelength red-shifted by 2.5~GHz from the $C_9$ transition and delivering a beam with a diameter of approximately 1.2~mm and a power of 1~mW.

The optical rotation angle of the linearly polarized probe beam is proportional to the transverse polarization of metastable atoms~\cite{PhysRevApplied.10.014002}. Thus, the observed FID signals, $S_\mathrm{FID}(t)=\sum_{i=1}^{8} S_i(0) \sin(\omega_i t) e^{-\Gamma_i t}$, arise from a linear superposition of eight eigenmodes supported by the full model, where $S_i(0)$, $\omega_i$, and $\Gamma_i$ denote the initial amplitude, precession frequency and relaxation rate of the $i$-th mode, respectively. The extreme brevity of the fast mode lifetimes limits their influence to the initial transient, with the subsequent evolution primarily governed by two degenerate slow modes. Consequently, the FID signals can be described by a single-mode approximation, $S_\mathrm{FID}(t)=S_\mathrm{s}(0)\sin(\omega_\mathrm{s} t)e^{-\Gamma_\mathrm{s} t}$, where the subscript denotes the slow mode.

We extract the procession frequency and relaxation rate from the $^3\mathrm{He}$ FID signals using polarimetric detection~\cite{PhysRevApplied.22.014084}. The external static field and circularly polarized pump light are aligned along the $x$-axis, while a transverse polarization is induced by an RF pulse applied along the $y$-axis and detected by a probe beam propagating in the same direction. The FID measurement procedure follows three sequential steps, depicted in Fig.~\ref{Figure3}(b). First, the pump light polarizes the metastable atoms, enabling ground-state nuclear spins to be polarized via ME collisions. Second, once the pump light has been switched off, a weak transverse RF pulse is applied to slightly tilt the collective nuclear spin away from the $x$-axis~\cite{Shaham2022}. We choose a tilt angle of approximately $7^\circ$ for our experiment to ensure that the transverse polarization remained within the fluctuation regime required for the linearization method adopted in the theoretical model. Third, the Faraday rotation signals of a weak linearly polarized probe light are measured after switching off the RF pulse. 

\section{Results and discussion}\label{four}
We analyze the real (relaxation) and imaginary (frequency) components of the slow mode across various nuclear polarization conditions. Initial measurements are conducted at $P = 0.08$ and used to fit the analytical solutions of the simplified model to calibrate ME-related parameters. We choose a low polarization to suppress tensorial contributions and ensure accurate fitting. Experiments at higher polarization such as $P = 0.58$ reveal alignment tensor effects in ME-coupled $^3\mathrm{He}$ and simultaneously validate the full model using calibrated parameters. Throughout the study, the nuclear polarizations are measured using the optical absorption method~\cite{refId0}.

\subsection{ME-induced relaxation}\label{fourA}
The FID signals are characterized by a decay rate $\Gamma_2=\Gamma_\mathrm{g}+\Gamma_\mathrm{ME}$, which is associated with two underlying physical processes. The first component, $\Gamma_\mathrm{g}$, is the intrinsic ground-state relaxation, which predominantly originates from wall collisions and magnetic-field inhomogeneities. The second, $\Gamma_\mathrm{ME}$, represents ME-induced relaxation, for which an analytical expression can be derived within the simplified model given by (\ref{MEC_relaxation}) in \ref{appendixC}. 
With a measured value of $\Gamma_\mathrm{g} \approx 0.05~\mathrm{s^{-1}}$, the intrinsic ground-state relaxation remains minuscule relative to the ME-induced contribution. Thus, across the magnetic field range of 1.0~$\mu\mathrm{T}$ to 38.6~$\mu\mathrm{T}$, the FID decay is dominated by ME collisions. The experimental data at $P = 0.08$ are fitted to (\ref{MEC_relaxation}), yielding calibrated ME collision rates of $1/T_e = 7.01~\mathrm{s^{-1}}$ and $1/\tau_e = 3.15\times10^{6}~\mathrm{s^{-1}}$.
The above parameters are treated as intrinsic to the ground-state--metastable hybrid $^3\mathrm{He}$ ensembles.

\begin{figure}[!t]
\centering
\includegraphics[scale=0.85]{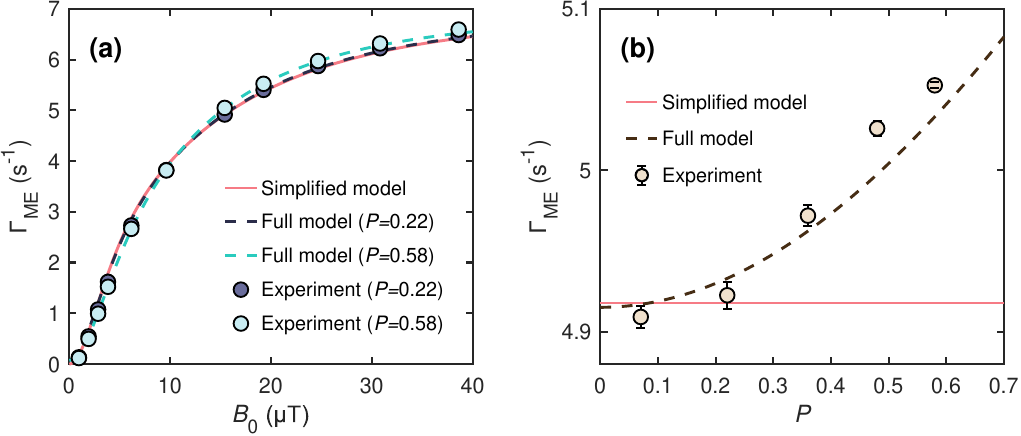}
\caption{(a) Magnetic field dependence and (b) nuclear polarization dependence of ME-induced relaxation. Simulation parameters $1/T_e = 7.01~\mathrm{s^{-1}}$ and $1/\tau_e = 3.15\times10^{6}~\mathrm{s^{-1}}$ are used throughout. In panel (b), the magnetic field is fixed at $B_0 = 15~\mu\mathrm{T}$.} 
\label{Figure4}
\end{figure}

As illustrated by the red solid line in Fig.~\ref{Figure4}(a), the simplified model predicts a monotonic increase of the ME-induced relaxation rate with field strength, asymptotically approaching $1/T_e$ in the high field. By incorporating the fitted parameters into the full model, the magnetic-field response of the ME-induced relaxation can be further evaluated at different nuclear polarizations. The dark and light dashed curves in Fig.~\ref{Figure4}(a) correspond to numerical solutions at $P = 0.22$ and $P = 0.58$, respectively. As expected, given that the condition at $P = 0.22$ remains within the applicable range of the simplified framework discussed in~\cite{DupontRoc1973}, the dark dashed line shows negligible departure from the red solid curve. In contrast, the onset of a pronounced deviation reflects the growing contribution of the alignment tensor at elevated polarization $P = 0.58$.

Validation of the full model is initially performed at pump powers of 25~mW and 385~mW, selected to match the polarization parameters adopted for the theoretical simulations presented in Fig.~\ref{Figure4}(a). As indicated by the dark and light dots, the results agree well with the full model predictions. Notably, in our experimental setup, the highly polarized $^3\mathrm{He}$ exhibits a lower ME-induced relaxation rate when the external field is below $10~\mu\mathrm{T}$; this trend reverses at higher field strengths. Such crossover behavior is consistently observed in both measurements and simulations, indicating that the introduction of the alignment tensor affects the coherence of ME-coupled $^3\mathrm{He}$ ensembles. As the error bars are smaller than the marker size, they are not shown in Fig.~\ref{Figure4}(a).

To further elucidate the dependence of ME-induced relaxation on the alignment tensor, we fixed the external magnetic field at $15~\mu\mathrm{T}$ while varying the polarization by continuously adjusting the pump laser power. In Fig.~\ref{Figure4}(b), the dashed line represents the full model theory, with dots denoting experiments with error bars derived from three independent repetitions. Within the range of polarizations considered, the experimental results agree well with the full model, confirming the validity of the theoretical description of polarization-dependent phenomena. For comparison, we also present the solution of the simplified model, computed from (\ref{MEC_relaxation}) and shown as the solid line, which is independent of polarization. In summary, the alignment tensor effects on ME-induced relaxation are observed by focusing on the real part of the slow mode. Meanwhile, we experimentally validate the full model in capturing the relaxation characteristics of highly polarized $^3\mathrm{He}$ under RF discharge.

\subsection{ME-induced frequency shift}\label{fourB}
\begin{figure}[!t]
\centering
\includegraphics[scale=0.85]{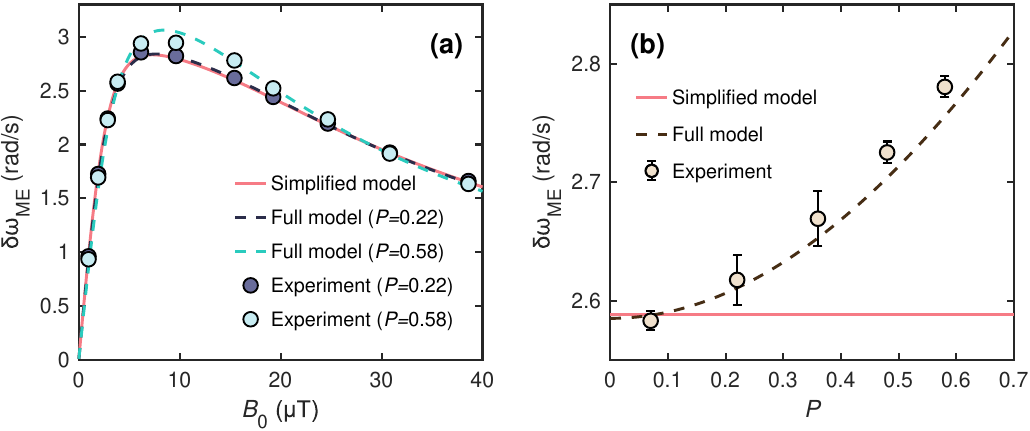}
\caption{(a) Magnetic field dependence and (b) nuclear polarization dependence of ME-induced frequency shift. Simulation parameters are the same as those of Fig.~\ref{Figure4}.} 
\label{Figure5}
\end{figure}

The oscillation frequency $\omega_\mathrm{s}$ of the FID signals primarily comprises the Larmor frequency $\omega_\mathrm{n}$ under a static magnetic field and the ME-induced frequency shift $\delta\omega_\mathrm{ME}$. Details of the determination of the static magnetic field strength from the FID signals are provided in \ref{appendixD}. Based on the ME collision rates calibrated in the previous section, Fig.~\ref{Figure5}(a) presents the magnetic-field dependence of the ME-induced frequency shift. The red solid curve shows the simplified model given by (\ref{delta_MEC}), in which the orientation predominantly determines the frequency shift. It exhibits a non-monotonic response to the magnetic field, rising to a maximum in the weak-field regime and declining thereafter. The full model predicts a gradual increase in the peak amplitudes of the shift curves due to the accumulation of alignment tensors, equivalent to the introduction of potential systematic errors exceeding 10~nT. 
Unlike the ME-induced relaxation, the shift curves for different polarizations feature two intersection points at magnetic fields around 4~$\mu$T and 30~$\mu$T. These intersections indicate that the ME-induced frequency shift is insensitive to tensor effects at specific field strengths. For the same reason as in Fig.~\ref{Figure4}(a), error bars are omitted in Fig.~\ref{Figure5}(a).
Following the analysis presented in the previous section, we further illustrate the dependence of the ME-induced frequency shift on nuclear polarization at a fixed field of $B_0=15~\mu\mathrm{T}$, which is shown in Fig.~\ref{Figure5}(b). Despite potential fluctuations introduced by the instability of the RF discharge and the external field, the observations are consistent with the expected behavior. These results indicate that the ME-induced frequency shift introduces systematic errors into $^3\mathrm{He}$ high-accuracy magnetometry that vary with the nuclear polarization, rendering the simplified model inadequate for accurate magnetic field reconstruction. Our method thus provides a numerical framework for calibrating higher-order polarization effects in MEOP-based $^3\mathrm{He}$ magnetometers, supporting the development of high-accuracy magnetic-field measurement~\cite{3He_LYF}.

We also evaluate the frequency shift induced by magnetic-field inhomogeneities. In our experimental setup, the cylindrical cell is subjected to magnetic field gradients $\nabla B_{y,z} < 10~\mathrm{nT/cm}$ and $\nabla B_x < 40~\mathrm{nT/cm}$, which corresponds to a negligible shift of $\delta\omega_\mathrm{grad} < 3.8 \times 10^{-3}~\mathrm{rad/s}$~\cite{PhysRevA.84.053411}. The gradient-induced shift is three orders of magnitude smaller than the ME-induced shift over the experimental magnetic field range. Therefore, under the present conditions, the field gradients exert no appreciable influence on the outcomes. 
Overall, by studying the imaginary parts of the slow mode, we observe the alignment tensor effects on the ME-induced frequency shift and provide a final validation of the full model.

\section{Conclusion}\label{five}
In this work, we employ a mean-field approximation to formulate a linearized model of ME collisions for the investigation of alignment tensor effects in the metastable $F=3/2$ manifold of ME-coupled $^3\mathrm{He}$.
Compared with conventional approaches, our method captures rank-2 tensorial contributions and uncovers previously overlooked polarization-dependent phenomena.
Specifically, alignment tensor effects on ME-induced relaxation and frequency shift are observed via FID measurements at high nuclear polarization across a wide range of external magnetic field strengths.
Eigenspectrum analysis of the ME-coupling structure shows good agreement between experimental results and theoretical predictions, further confirming the validity of the theory.
This work advances the fundamental understanding of nuclear spin dynamics in highly polarized $^3\mathrm{He}$ prepared by MEOP. 
It also provides a framework for treating nT-level systematic errors due to alignment tensors in high-accuracy magnetometry~\cite{3He_LYF}, and for designing optimal protocols for preparing nuclear spin-squeezed states~\cite{PhysRevLett.127.013601}.

%
%

\ack{This work was supported by the National Natural Science Foundation of China (Grant No.~62375002).}



\data{The data that support the findings of this article are not publicly available. The data are available from the authors upon reasonable request.}


\appendix
\renewcommand{\thesection}{Appendix \Alph{section}}
\section{Alignment tensor evolution in the irreducible spherical tensor basis}\label{appendixA}
The ME collisions equations of motion for the alignment tensor $Q_{\alpha\beta}$ are formulated in Cartesian coordinates, as detailed in (\ref{MEC}).
Using the transformation given in (\ref{basis_relation}), the ME dynamics can be recast into the irreducible spherical representation,
\begin{equation}
    \begin{aligned}
        \left.\frac{\mathrm{d}}{\mathrm{d}t} \langle \Re T_2^2 \rangle\right|_\text{ME} &= -\frac{2}{3\tau_e} \langle \Re T_2^2 \rangle + \frac{1}{\sqrt{3}\tau_e} \frac{1}{N_\mathrm{t}} \left( \langle I_x \rangle \langle\Sigma_x \rangle - \langle I_y \rangle \langle \Sigma_y \rangle \right),\\
        \left.\frac{\mathrm{d}}{\mathrm{d}t} \langle \Im T_2^2 \rangle\right|_\text{ME} &= -\frac{2}{3\tau_e} \langle \Im T_2^2 \rangle + \frac{1}{\sqrt{3}\tau_e} \frac{1}{N_\mathrm{t}} \left( \langle I_x \rangle \langle \Sigma_y \rangle + \langle I_y \rangle \langle \Sigma_x \rangle \right),\\
        \left.\frac{\mathrm{d}}{\mathrm{d}t} \langle \Re T_1^2 \rangle\right|_\text{ME} &= -\frac{2}{3\tau_e} \langle \Re T_1^2 \rangle - \frac{1}{\sqrt{3}\tau_e} \frac{1}{N_\mathrm{t}} \left( \langle I_x \rangle \langle \Sigma_z \rangle + \langle I_z \rangle \langle \Sigma_x \rangle \right),\\
        \left.\frac{\mathrm{d}}{\mathrm{d}t} \langle \Im T_1^2 \rangle\right|_\text{ME} &= -\frac{2}{3\tau_e} \langle \Im T_1^2 \rangle - \frac{1}{\sqrt{3}\tau_e} \frac{1}{N_\mathrm{t}} \left( \langle I_y \rangle \langle \Sigma_z \rangle + \langle I_z \rangle \langle \Sigma_y \rangle \right),\\
         \left.\frac{\mathrm{d}}{\mathrm{d}t} \langle T_0^2 \rangle\right|_\text{ME} &= -\frac{2}{3\tau_e} \langle T_0^2 \rangle + \frac{1}{3\tau_e} \frac{1}{N_\mathrm{t}} \left( 3 \langle I_z \rangle \langle \Sigma_z \rangle - \langle \mathbf{I} \rangle \cdot \langle \boldsymbol{\Sigma} \rangle \right).
    \end{aligned}
\end{equation}
To describe the evolution of the alignment tensor in an external magnetic field, we first introduce the commutation relations required for the derivation~\cite{PhysRevA.76.033830}, 
\begin{equation}
    \begin{aligned}
         &\left[F_{3/2,z},T^2_q\right]=qT^2_q,\\
         &\left[F_{3/2,\pm},T^2_q\right]=\sqrt{\left(2\pm q+1\right)\left(2\mp q\right)}T^2_{q\pm 1},
    \end{aligned}
\end{equation}
while $\mathbf{F}_{1/2}$ and $\mathbf{I}$ commute with $T_q^2$. Using the Hamiltonian given in (\ref{Hamiltonian}) and the Heisenberg equation of motion, one arrives at the corresponding evolutions of the real and imaginary parts of $T_q^2$,
\begin{equation}
    \begin{aligned}
        \left.\frac{\mathrm{d}}{\mathrm dt}\Re T_2^2\right |_B&=\gamma_{3/2}\left( B_{0,x}\Im T_1^2+ B_{0,y}\Re T_1^2+2 B_{0,z}\Im T_2^2\right),\\
        \left.\frac{\mathrm{d}}{\mathrm dt}\Im T_2^2\right |_B&=\gamma_{3/2}\left(- B_{0,x}\Re T_1^2+ B_{0,y}\Im T_1^2-2 B_{0,z}\Re T_2^2\right),\\
        \left.\frac{\mathrm{d}}{\mathrm dt}\Re T_1^2\right |_B&=\gamma _{3/2}\left[B_{0,x}\Im T_2^2- B_{0,y}\left(\Re T_2^2-\sqrt{3} T_0^2\right)+ B_{0,z}\Im T_1^2\right],\\
         \left.\frac{\mathrm{d}}{\mathrm dt}\Im T_1^2\right |_B&=\gamma_{3/2}\left[ -B_{0,x}\left(\sqrt{3} T_0^2+\Re T_2^2\right)- B_{0,y}\Im T_2^2-B_{0,z}\Re T_1^2\right],\\
         \left.\frac{\mathrm{d}}{\mathrm dt} T_0^2\right |_B&=\sqrt{3}\gamma_{3/2}\left(B_{0,x}\Im T_1^2-B_{0,y}\Re T_1^2\right).     
    \end{aligned}
\end{equation}

\section{Longitudinal coupling matrix}\label{appendixB} 
This appendix presents the explicit form of the ME coupling matrix within the longitudinal subspace:
\begin{equation}
    \renewcommand{\arraystretch}{2} 
    \mathbf{M}_{11} =
    \begin{pmatrix}
        \begin{array}{@{\hspace{1em}}c @{\hspace{1em}} 
               c @{\hspace{1em}} 
               c @{\hspace{1em}} 
               c @{\hspace{1em}} 
               c @{\hspace{1em}} 
               c @{\hspace{1em}}}
            -\frac{1}{T_e} & \frac{1}{3\tau_e} & -\frac{1}{3\tau_e} & 0 & 0 & 0\\
            \frac{2(5+3P^2)}{3T_e(3+P^2)} & -\frac{4}{9\tau_e} & \frac{10}{9\tau_e} & -\frac{P}{9\tau_e} & 0 & \frac{P}{3\sqrt{3}\tau_e}\\
            -\frac{1+3P^2}{3T_e(3+P^2)} & \frac{1}{9\tau_e} & -\frac{7}{9\tau_e} & \frac{P}{9\tau_e} & 0 & -\frac{P}{3\sqrt{3}\tau_e}\\
            -\frac{4P}{3T_e(3+P^2)} & -\frac{P}{9\tau_e} & -\frac{2P}{9\tau_e} & -\frac{2}{3\tau_e} & -\sqrt{3}\omega_{3/2} & 0\\
            0 & 0 & 0 & \sqrt{3}\omega_{3/2} & -\frac{2}{3\tau_e} & \omega_{3/2}\\
            \frac{4P}{\sqrt{3}T_e(3+P^2)} & \frac{P}{3\sqrt{3}\tau_e} & \frac{2P}{3\sqrt{3}\tau_e} & 0 & -\omega_{3/2} & -\frac{2}{3\tau_e}
        \end{array}
    \end{pmatrix}.
\end{equation}

\section{Analytical solutions of the simplified model}\label{appendixC}
In the simplified model of ME collisions, which considers only the orientation of $^3\mathrm{He}$ ensembles, the dynamical equations for the atomic variables under an external magnetic field can be written as:
\begin{equation}
    \begin{aligned}
        \frac{\mathrm{d}}{\mathrm{d}t}\langle\mathbf{I}\rangle
        &=- \frac{1}{T_e}\langle\mathbf{I}\rangle+\frac{1}{3T_e}\frac{N_\mathrm{t}}{n_\mathrm{t}}\left(\langle\mathbf{F}_{3/2}\rangle-\langle\mathbf{F}_{1/2}\rangle\right)
        +\gamma_\mathrm{n}\langle\mathbf{I}\rangle\times\mathbf{B_0},\\
        \frac{\mathrm{d}}{\mathrm{d}t}\langle\mathbf{F}_{3/2}\rangle
        &=-\frac{4}{9\tau_e}\langle\mathbf{F}_{3/2}\rangle+\frac{10}{9\tau_e}\langle\mathbf{F}_{1/2}\rangle+\frac{10}{9\tau_e}\frac{n_\mathrm{t}}{N_\mathrm{t}}\langle\mathbf{I}\rangle
        +\gamma_{3/2}\langle\mathbf{F}_{3/2}\rangle\times\mathbf{B_0},\\
        \frac{\mathrm{d}}{\mathrm{d}t}\langle\mathbf{F}_{1/2}\rangle
        &=-\frac{7}{9\tau_e}\langle\mathbf{F}_{1/2}\rangle+\frac{1}{9\tau_e}\langle\mathbf{F}_{3/2}\rangle-\frac{1}{9\tau_e}\frac{n_\mathrm{t}}{N_\mathrm{t}}\langle\mathbf{I}\rangle
        +\gamma_{1/2}\langle\mathbf{F}_{1/2}\rangle\times\mathbf{B_0}.
    \end{aligned}
\end{equation}
This allows Ref.~\cite{DupontRoc1973} to provide analytical expressions that describe the ME-induced relaxation and frequency shift:
\begin{equation}
    \Gamma_\mathrm{ME}
    =\frac{1}{T_e}\left[1-\frac{4}{9}\frac{1}{1+\left(\gamma_\mathrm{m}B_0\tau_e\right)^2}-\frac{5}{9}\frac{1}{1+16\left(\gamma_\mathrm{m}B_0\tau_e\right)^2}\right],\label{MEC_relaxation}
\end{equation}
\begin{equation}
    \delta\omega_\mathrm{ME}=-\frac{4\gamma_\mathrm{m}B_0\tau_e}{9T_e}\left[\frac{1}{1+\left(\gamma_\mathrm{m}B_0\tau_e\right)^2}+\frac{5}{1+16\left(\gamma_\mathrm{m}B_0\tau_e\right)^2}\right].\label{delta_MEC}
\end{equation}

\section{Determination of the static magnetic field strength}\label{appendixD}
The oscillation frequency $\omega_\mathrm{s}$ of the FID signals consists of the Larmor frequency $\omega_\mathrm{n}$ and ME-induced frequency shift $\delta\omega_\mathrm{ME}$. For sufficiently low nuclear polarization, the contribution of the alignment tensor becomes negligible, $\delta\omega_\mathrm{ME}$ is determined solely by the atomic orientation and admits the analytical form expressed as Eq.~(\ref{delta_MEC}). Thus, we have
\begin{equation}
    \left| \omega_\mathrm{s} \right|=\left|\gamma_\mathrm{n}B_0+\frac{4\gamma_\mathrm{m}B_0\tau_e}{9T_e}\left[\frac{1}{1+\left(\gamma_\mathrm{m}B_0\tau_e\right)^2}+\frac{5}{1+16\left(\gamma_\mathrm{m}B_0\tau_e\right)^2}\right]\right|.\label{omega_s}
\end{equation}
The absolute value in Eq.~(\ref{omega_s}) arises from the frequency extraction procedure, since the oscillation frequency $\omega_\mathrm{s}$ obtained from the FID signals is always positive in the experiment. We calibrate the magnetic field under weak optical-pumping conditions, with the $^3$He nuclear polarization maintained below 0.1. With the tensorial contribution being negligible under this condition, the measured FID oscillation frequency $\omega_\mathrm{s}$ can be described by Eq.~(\ref{omega_s}), which is equivalent to a quintic equation in magnetic field strength $B_0$ and can therefore be solved numerically. Assuming that $B_0$ remained unchanged during the subsequent measurements, the observed variation in $\omega_\mathrm{s}$ at higher nuclear polarizations can be primarily ascribed to changes in the ME-induced frequency shift $\delta\omega_\mathrm{ME}$.


\end{document}